\documentstyle[psfig,conf_iap,10pt]{article}
\def\h{~h$^{-1}$ Mpc~}

\def\hmentre{~h$^{3}$ Mpc$^{-3}$~}
\def\lsole{~L$_{\odot}$~}
\def\o2{[OII]~$\lambda 3727$}

\def\hmagp{$+5\log h$~}

\begin{document}
\heading{%
%
The luminosity function and mean galaxy density from the ESP galaxy 
redshift survey
%
} 
\par\medskip\noindent
\author{%
E.Zucca$^{1,2}$,
G.Zamorani$^{1,2}$,
G.Vettolani$^{2}$,
A.Cappi$^{1}$,
R.Merighi$^{1}$,
M.Mignoli$^{1}$,
\\
G.M.Stirpe$^{1}$,
H.MacGillivray$^{3}$,
C.Collins$^{4}$,
C.Balkowski$^{5}$,
V.Cayatte$^{5}$,
\\
S.Maurogordato$^{6,5}$,
D.Proust$^{5}$,
G.Chincarini$^{7,8}$,
L.Guzzo$^{7}$,
D.Maccagni$^{9}$,
\\
R.Scaramella$^{10}$,
A.Blanchard$^{11}$,
M.Ramella$^{12}$
}
\address{%
Osservatorio Astronomico di Bologna, Bologna, Italy
}
\address{%
Istituto di Radioastronomia del CNR, Bologna, Italy
}
\address{%
Royal Observatory Edinburgh, Edinburgh, United Kingdom
}
\address{%
Astrophysics Research Institute, Liverpool, United Kingdom
}
\address{%
DAEC, Observatoire de Paris, Meudon, France
}
\address{%
CERGA, Observatoire de la C\^ote d'Azur, Nice, France
}
\address{%
Osservatorio Astronomico di Brera, Merate (LC), Italy
}
\address{%
Universit\`a degli Studi di Milano,Milano, Italy
}
\address{%
Istituto di Fisica Cosmica e Tecnologie Relative, Milano, Italy
}
\address{%
Osservatorio Astronomico di Roma, Monteporzio Catone (RM), Italy
}
\address{%
Universit\'e L. Pasteur, Observatoire Astronomique, Strasbourg, France
}
\address{%
Osservatorio Astronomico di Trieste, Trieste, Italy
}
\section{The ESO Slice Project (ESP)}
The ESO Slice Project (ESP) \cite{esp1} \cite{esp3} galaxy redshift 
survey is intermediate between shallow, wide angle samples and very
deep, monodimensional pencil beams.
It extends over a strip of $\alpha \times \delta = 22^o \times 
1^o$, plus a nearby area of $5^o \times 1^o$, five degrees west of the main 
strip, in the South Galactic Pole region.
The right ascension limits are $ 22^{h} 30^m$ and $ 01^{h} 20^m $, at a mean 
declination of $ -40^o$ (1950); the covered solid angle is $\sim 23$ sq. deg. 
The limiting magnitude of the survey is $b_J \leq 19.4$, which optimizes the
number of fibers of the spectrograph and corresponds to an effective depth
of $z \sim 0.16$.  
\\
We observed a total of 4044 spectra, corresponding to $\sim 90\%$ of the 
starting photometric sample, which contains 4487 objects.
Among the observed spectra, 493 objects turned out to be stars ($\sim 12\%$
of the observed objects) and 208 were not useful to derive redshift
determinations.
Therefore we obtained a total of 3342 galaxies ($+$ a QSO) with reliable 
redshifts: about half of these galaxies present spectra with emission lines.
The wedge diagram of the observed galaxies is shown in Fig.1.
\\
We paid particular attention to the redshift quality and we applied several
checks to the data, using also multiple observations of $\sim 200$ galaxies
and $\sim 750$ galaxies for which the redshift from both absorption and
emission line is available \cite{esp3} \cite{esp4}. Among the various analyses
we applied to the ESP sample,    
here we summarize our results on the luminosity function \cite{esp2}.
\begin{figure} 
\centerline{\vbox{
\psfig{figure=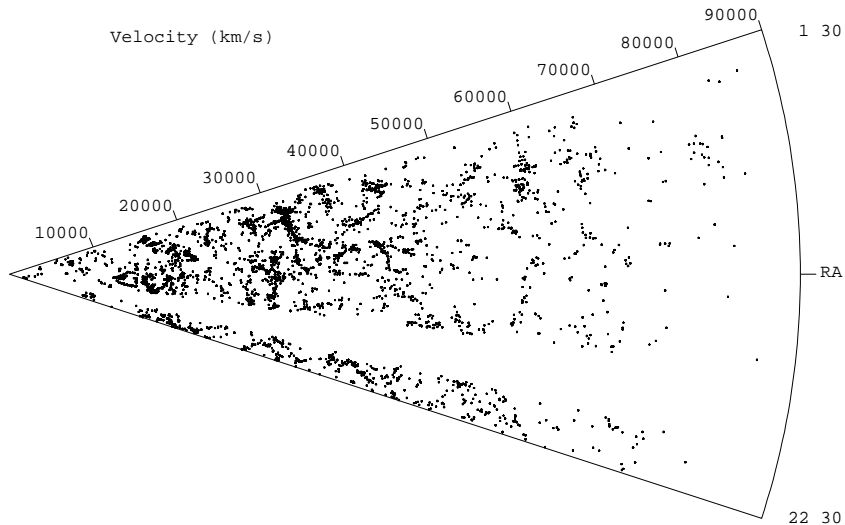,height=7.cm}
}}
\caption[]{Wedge diagram of the 3342 ESP galaxies. ~~~~~~~~~~~~~~~~~~~~~~~~~~~~~~~~
}
\end{figure}
\begin{figure}[t]
\centerline{{
\psfig{figure=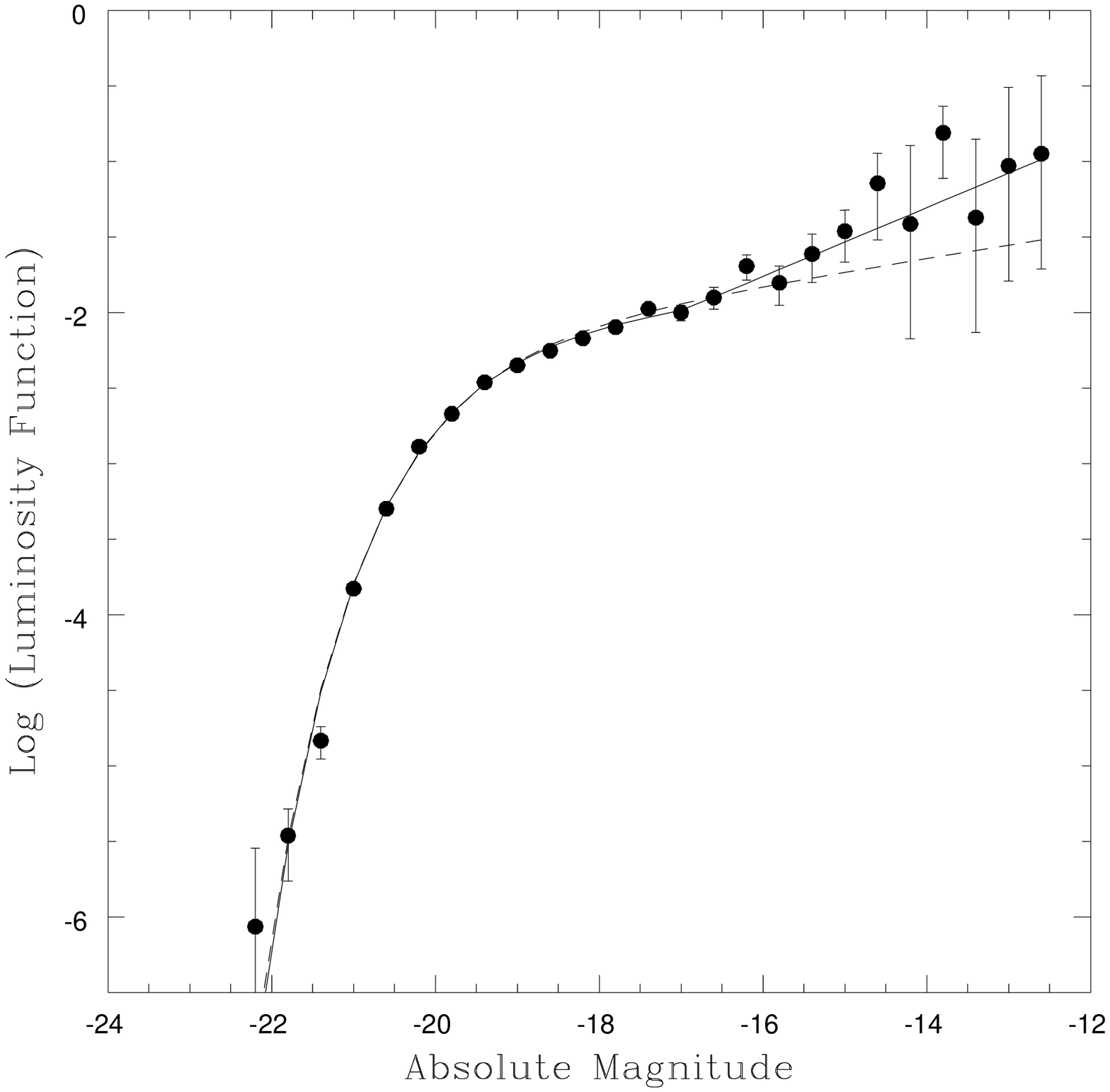,width=0.5\hsize}
\psfig{figure=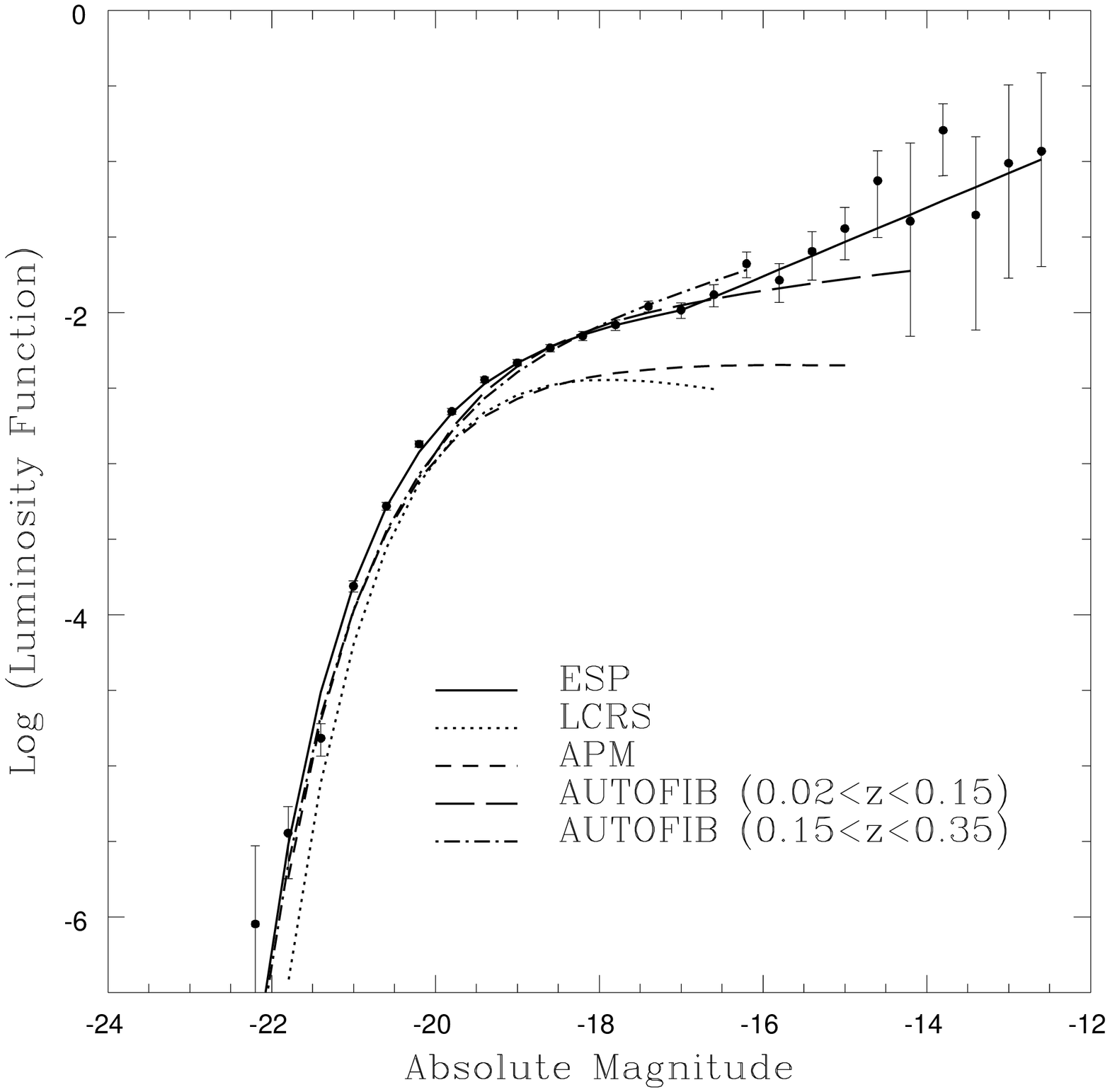,width=0.5\hsize}
}}
\caption[]{a) Normalized luminosity function for 3342 ESP galaxies brighter
than $M_{b_J}=-12.4$ \hmagp. The solid circles are computed with a modified 
version of the C--method,  while the fits are obtained with the STY method. 
Dashed line: single Schechter function; solid line: Schechter function and 
power law. b) Comparison of the ESP luminosity function with previous results.
}
\end{figure}
\begin{figure}
\centerline{{
\psfig{figure=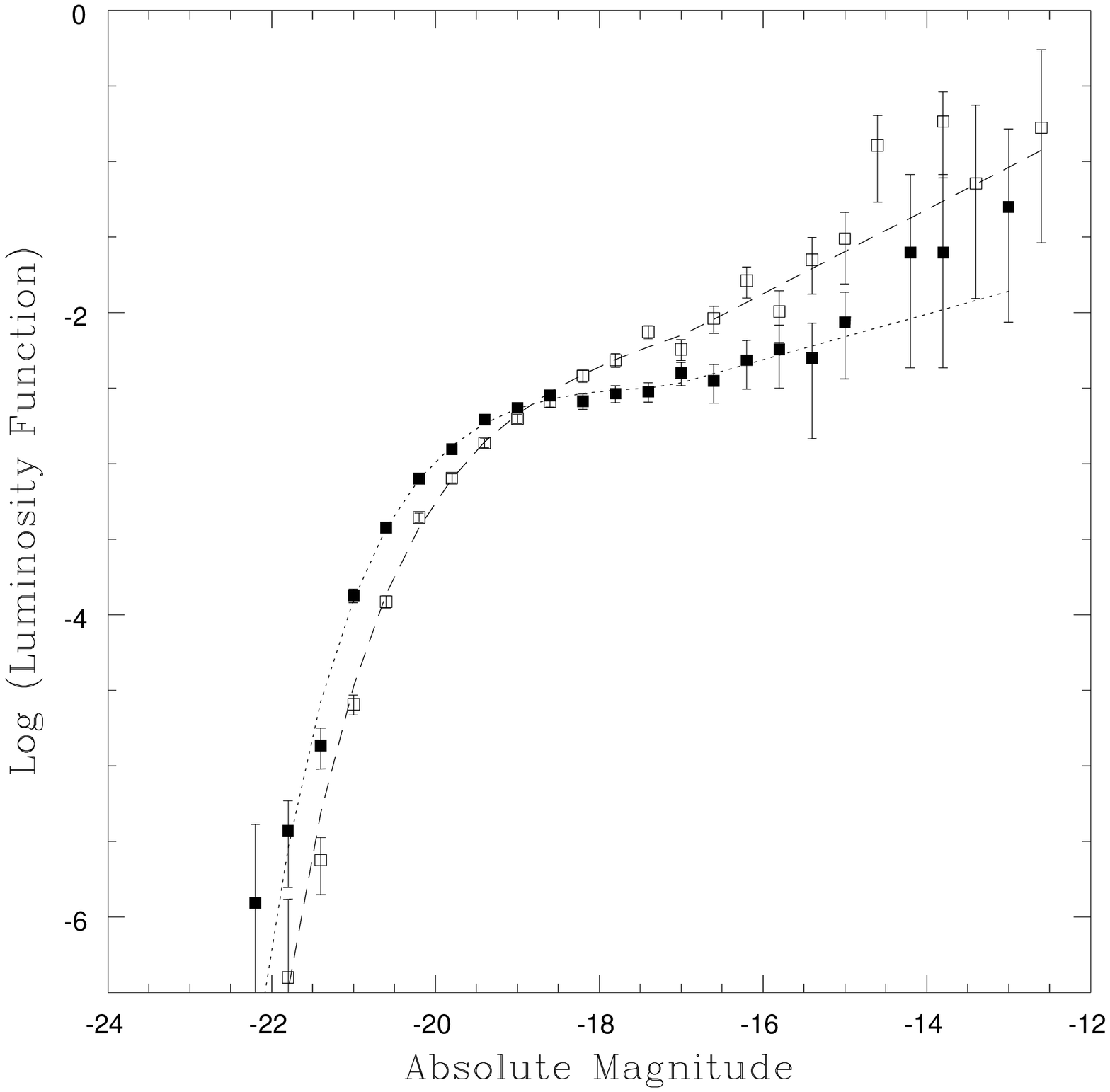,width=0.5\hsize}
\psfig{figure=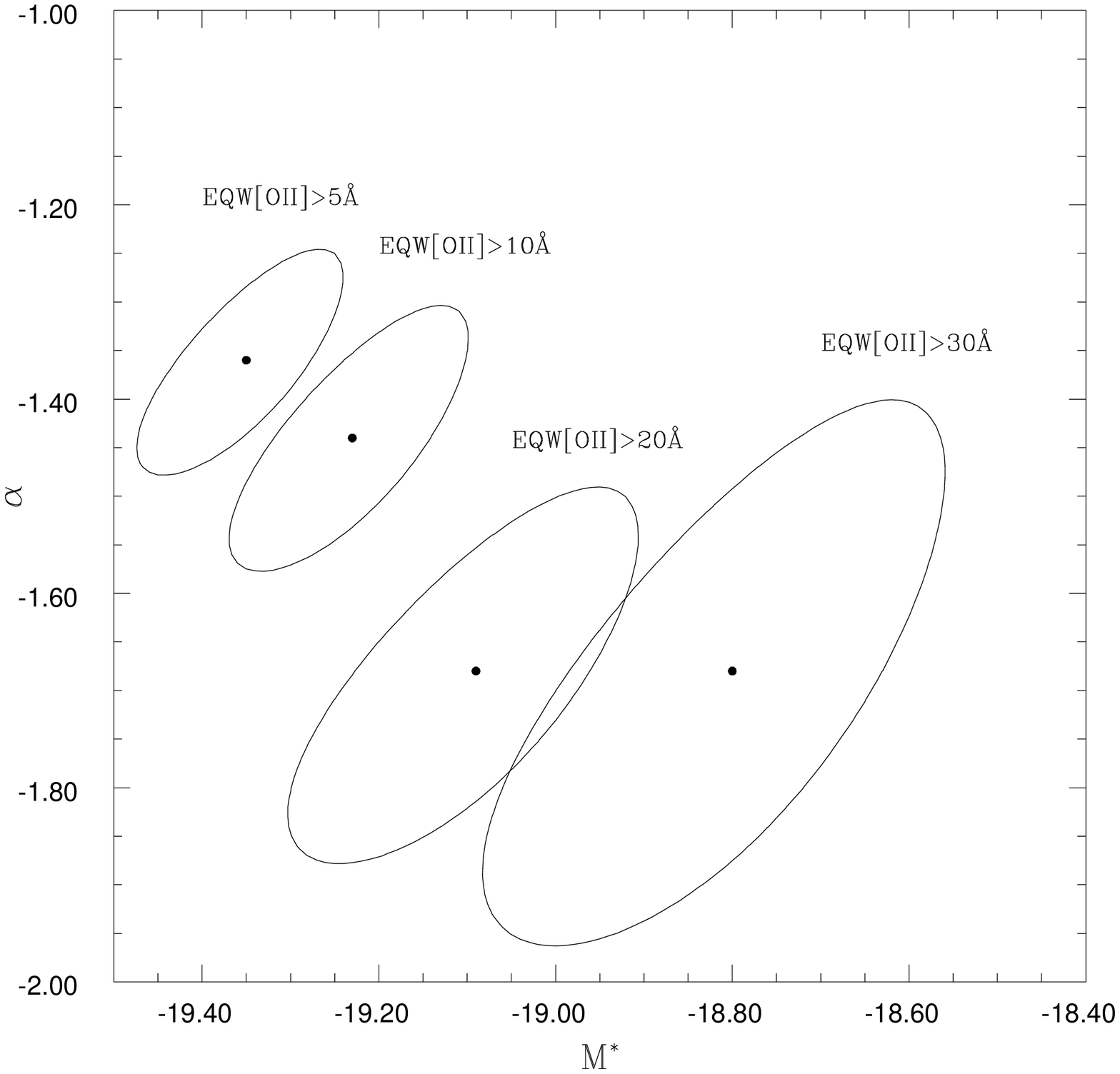,width=0.5\hsize}
}}
\caption[]{a) The same as Fig.2a but for galaxies with (open squares and dashed 
line) and without (filled squares and dotted line) emission lines. 
\\
b) Confidence ellipses for the parameters $\alpha$ and $M^*$ of the luminosity
functions of galaxy subsamples with different thresholds in [OII] equivalent
width. 
}
\end{figure}
\begin{figure}
\centerline{{
\psfig{figure=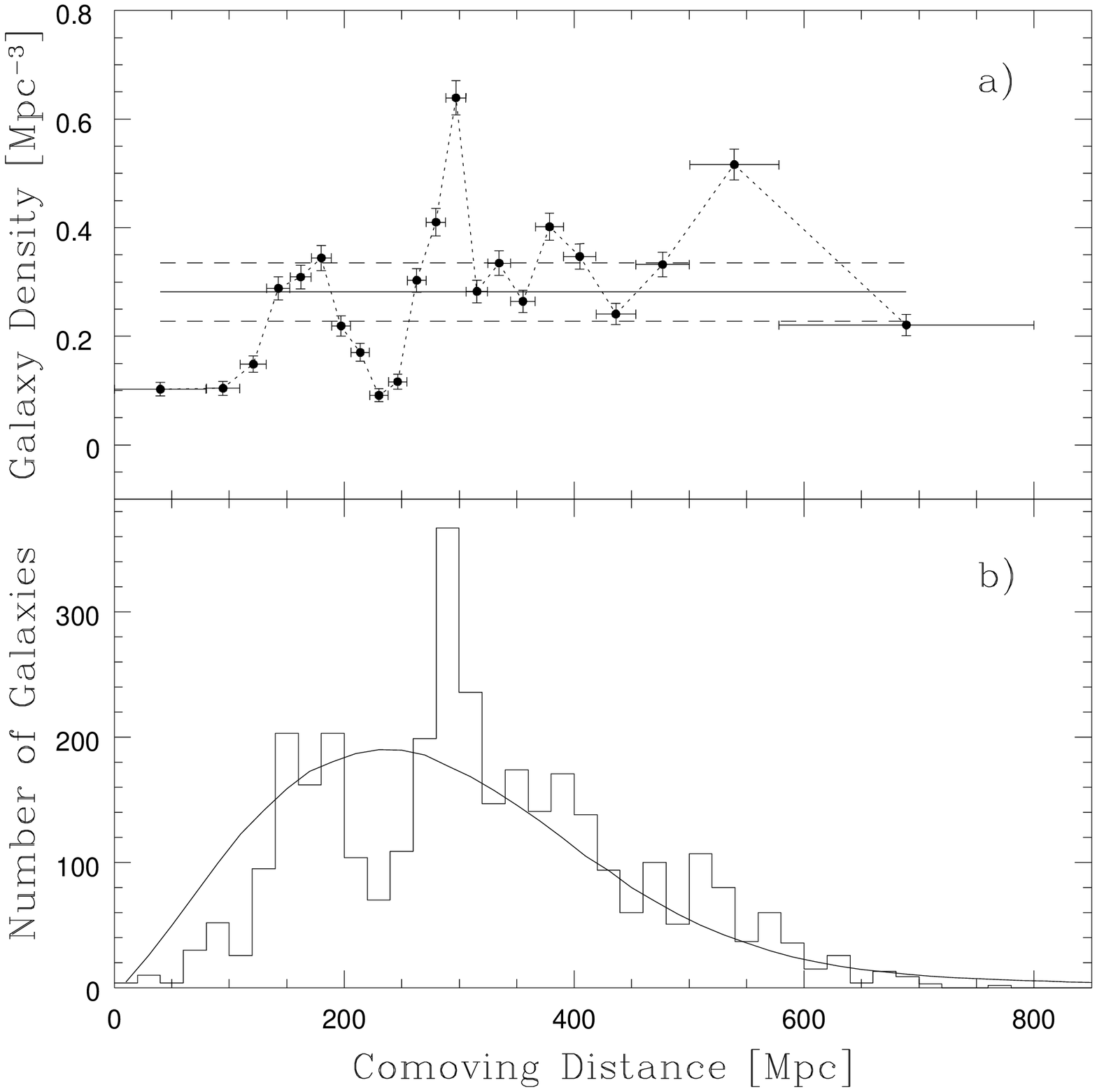,width=0.5\hsize}
\psfig{figure=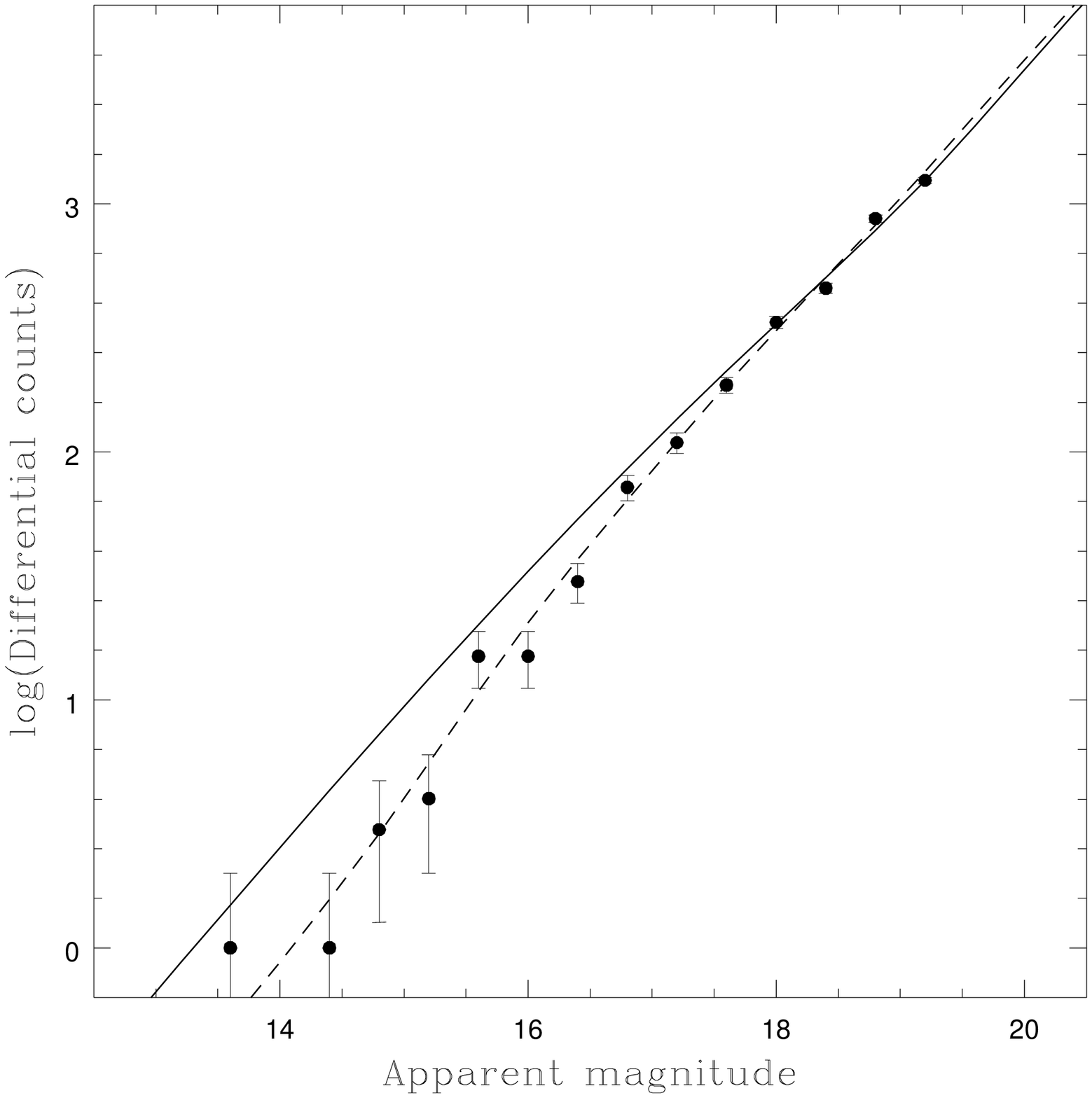,width=0.5\hsize}
}}
\caption[]{a) Galaxy density as a function of the comoving distance, with
superimposed the mean value $\bar n \pm 1\sigma$ (solid and dashed lines),
for galaxies with $M_{b_J}\le-12.4$ \hmagp (single Schechter function case). 
b) Comoving distance histogram of ESP galaxies: the solid line is the
distribution expected for a uniform sample.
c) Observed galaxy counts compared with those expected from the 
luminosity function (solid line); the same but taking into account the observed
radial density variations (dashed line).
}
\end{figure}
\section{The luminosity function and the mean galaxy density } 
Given its characteristics, the ESP sample can provide an accurate estimate of 
both the normalization and the faint end slope of the luminosity function in 
the local Universe. Our main results \cite{esp2} are the following:
\\
1) Although a Schechter function is an acceptable representation of the
luminosity function over the entire range of magnitudes ($M_{b_J}\le-12.4$
\hmagp), our data suggest the presence of a steepening of the luminosity 
function for $M_{b_J}\ge -17$ \hmagp (Fig.2a), well fitted by a 
power law with slope $\beta \sim -1.6$.
\\
2) The steepening at the faint end of the luminosity function is almost 
completely due to galaxies with emission lines: in fact, dividing our galaxies 
into two samples, i.e. galaxies with and without emission lines, we find 
significant differences in their luminosity functions. In particular, galaxies 
with emission lines show a significantly steeper slope and a fainter $M^*$
(Fig.3a). 
This trend depends on the [OII] equivalent width: when considering subsamples 
of galaxies at increasing thresholds of [OII] equivalent width, $\alpha$ 
becomes steeper and $M^*$ fainter (Fig.3b). 
\\
3) The amplitude and the $\alpha$ and $M^*$ parameters of our luminosity
function are in good agreement with those of the AUTOFIB redshift survey
\cite{auto}.
Viceversa, our amplitude is a factor $\sim 1.6$ higher, at $M\sim M^*$,
than that found for both the Stromlo-APM \cite{apm} and the Las Campanas 
\cite{lcrs} redshift surveys. Also the faint end 
slope of the luminosity function is significantly steeper for the ESP 
galaxies than that found in these two surveys (Fig.2b).
\\
4) The galaxy number density for $M_{b_J}\le -16$ \hmagp is well determined 
($\bar n = 0.08 \pm 0.015$ \hmentre). 
Its estimate for $M_{b_J}\le -12.4$ \hmagp is more 
uncertain, ranging from  $\bar n = 0.28$ \hmentre, in the case of a fit with a 
single Schechter function, to $\bar n = 0.54$ \hmentre, in the case of 
a fit with a Schechter function and a power law.
The corresponding blue luminosity densities are 
$\rho_{LUM}= (2.0, 2.2, 2.3) \times 10^8\ h$ \lsole Mpc$^{-3}$, respectively.
\\
5) We find evidence for a local under--density, extending up to a comoving
distance $\sim 140$ \h (Fig.4a and 4b). 
When the radial density variations observed in our
data are taken into account, our derived luminosity function reproduces
very well the observed counts for $b_J \le 19.4$, including the steeper than
Euclidean slope for $b_J \le 17.0$ (Fig.4c). If this under--density extends 
over a much larger solid angle than that covered by our survey, it could, at 
least partly, explain the low amplitude of the Stromlo-APM luminosity function.
\\
6) A similar explanation can not justify the significant difference in
amplitude between the ESP and the LCRS luminosity functions, because
the two samples cover essentially the same redshift range. One possibility,
which has however to be verified, is that a non negligible number of
galaxies are missing from the original CCD photometric catalog of the LCRS.
\vfill
\end{document}